\documentclass[sigconf]{acmart}

\usepackage{float}
\usepackage{booktabs}
\usepackage{amsmath}
\usepackage{pifont}
\usepackage{microtype}
\newcommand{\cmark}{\ding{51}}
\newcommand{\nobreakeven}{\textemdash}

\usepackage{graphicx}
\usepackage[dvipsnames]{xcolor}

\usepackage{algorithm}
\usepackage{algpseudocode}

\usepackage{adjustbox}
\usepackage{fancyvrb}

\usepackage{ragged2e}
\usepackage{etoolbox}

\AtEndEnvironment{algorithm}{\justifying}
\AtEndEnvironment{algorithm*}{\justifying}

\AfterEndEnvironment{verbatim}{\par\justifying\normalfont}

\AtBeginDocument{%
  }

\setcopyright{acmlicensed}
\copyrightyear{2025}
\acmYear{2025}
\acmDOI{XXXXXXX.XXXXXXX}
\acmConference[HPDC 2026]{xxxx}{July 13-16, 2026,}{Cleveland, OH, USA}
\acmISBN{978-1-4503-XXXX-X/2018/06}
\begin{document}

\title{Analyzing Persistent Alltoallv RMA Implementations for High-Performance MPI Communication\\}

\author{Evelyn Namugwanya}
\email{enamugwan42@tntech.edu}
%\orcid{0000-0000-0000-0000}
\affiliation{%
  \institution{Tennessee Tech University}
  \city{Cookeville}
  \state{Tennessee}
  \country{USA}
}
\begin{comment}
\author{Joseph Schuchart}
\email{Joseph.Schuchart@stonybrook.edu}
\affiliation{%
  \institution{Stony Brook University}
  \city{Stony Brook}
  \state{New York}
  \country{USA}
}
\author{Anthony Skjellum}
\email{askjellum@tntech.edu}
\affiliation{%
  \institution{Tennessee Tech University}
  \city{Cookeville}
  \state{Tennessee}
  \country{USA}
}
\end{comment}
% \author{}
% \affiliation{%
%   \institution{}
%   \city{}
%   \state{}
%   \country{}}
% \email{}

% \author{}
% \affiliation{%
%   \institution{}
%   \city{}
%   \state{}
%   \country{}}
% \email{}


\begin{abstract}
 
Collective communication operations such as \texttt{MPI\_Alltoallv} are central to many HPC applications, particularly those with irregular message sizes. We design, implement, and evaluate \emph{persistent} MPI RMA variants of \texttt{Alltoallv} based on fence and lock synchronization, separating a one-time initialization phase from per-iteration execution to enable reuse of communication metadata and window state across repeated epochs. 

Our Benchmarks tested on LLNL’s Dane supercomputer show that the fence-persistent variant consistently outperforms the non-persistent baseline for large message sizes, achieving up to 44\% reduction in runtime and improving scalability with increasing process counts; at 448 processes the runtime decreases from 2.49\,s to 1.54\,s (38\% faster). We further evaluate the algorithms under irregular sparse communication patterns and compare fence- and lock-based designs, including hierarchical extensions.

Message-size sweeps and a break-even model demonstrate that persistence provides immediate payoff for messages $\ge 32{,}768$~bytes, while smaller messages show limited benefit due to metadata amortization costs. These results indicate that persistent RMA \texttt{Alltoallv} is a practical approach for workloads with large messages, where removing repeated metadata processing leaves runtime dominated by data movement, as evidenced by the increasing time savings with message size, and they clarify the trade-offs between fence and lock synchronization on modern HPC systems.

\end{abstract}

\maketitle

\section{Introduction}

Efficient communication is key in high performance computing, especially as modern applications increasingly scale across thousands of processors and heterogeneous architectures. \texttt{MPI\_Alltoallv} is vital in parallel programs that involve irregular communication patterns, including applications in scientific computing, sparse linear algebra, and particle simulations. The \texttt{MPI\_Alltoallv} routine enables processes to exchange variable-sized messages, making it highly flexible but potentially expensive at scale due to its reliance on two-sided communication within implementations, as well as synchronization across all participating processes.

To address these scalability challenges, MPI provides Remote Memory Access (RMA) functionality, also known as one-sided communication. Unlike traditional point-to-point messaging, RMA enables a process to directly read from or write to the memory of a remote process without requiring the remote side’s active participation during the operation.

Despite its theoretical benefits, RMA remains underexplored for collective communication patterns like \texttt{Alltoallv}. Most prior research has focused on point-to-point communication in underlying implementations, leaving a gap in understanding how RMA-based collectives perform. The approach here is to layer an MPI collective communication over MPI RMA instead of implementing it in terms of MPI point-to-point operations, a widespread choice for the past 30+ years. This works for persistent mode, new in MPI-4\cite{mpi40}.

The main idea of this work is to introduce persistent RMA collectives, a capability not provided by the MPI standard. MPI defines persistent point-to-point operations and persistent two-sided collectives, but it does not provide persistence for RMA-based collectives. Existing RMA interfaces require window creation, datatype decoding, displacement exchange, and synchronization setup to be repeated on every invocation, even when the communication pattern is unchanged. Our work fills this gap by introducing a persistent RMA formulation of \texttt{MPI\_Alltoallv} that separates one-time initialization from per-iteration execution. Our proposed design caches RMA windows, remote displacements, datatype information, and communication schedules inside a reusable request object and exposes start/wait semantics for repeated RMA epochs.

This paper makes the following contributions:
\begin{itemize}
\raggedright
    \item We design and implement novel persistent RMA-based \texttt{Alltoallv} algorithms namely \texttt{Fence persistent} (collective synchronization), \texttt{Lock persistent} (passive target synchronization) and \texttt{Fence\_Hirarchy\_persistent} (collective synchronization with ordered MPI\_Puts), overlapping remote and local transfers.
    
    \item We design two benchmark frameworks to evaluate these RMA variants across message sizes, system scales, and regular and irregular communication patterns.
    
    \item We describe comprehensive experiments on \footnote{Open Systems on Lawrence Livermore National Laboratory (LLNL)}---Dane super computer and compare our implementations to the standard \texttt{MPI\_Alltoallv}.
  
    \item We present key findings:
    \begin{itemize}
        \item Persistent RMA variants outperform traditional \texttt{MPI\_Alltoallv} at large message sizes ($\ge 32{,}768$~bytes).
         \item Persistent RMA variants scale more effectively at high process counts, reducing runtime up to 38\% at 448 processes.
        \item Fence-based persistent RMA (\texttt{Fence persistent}) consistently performs better than lock-based persistent RMA (\texttt{Lock persistent}).
       % \item Traditional \texttt{Alltoallv} remains more efficient for small messages because of lower latency overheads.
    \end{itemize}
\end{itemize}

\subsection{The Role of Persistence}
All our RMA-based algorithms are designed with persistence, separating initialization 
from execution. The persistent request object stores buffer information, displacements, 
and synchronization routines, avoiding repeated window creation and metadata exchange. 
This persistence reduces overhead significantly when the same communication pattern is 
invoked repeatedly, as in iterative applications. 
By amortizing setup costs, persistence ensures better scalability and higher efficiency 
compared to non-persistent.

The benefit of persistence can be expressed through a break-even analysis. Let 
$T_{\text{nonpersistent}}$ denote the runtime per iteration of a standard 
(non-persistent) Alltoallv, $T_{\text{persistent}}$ the per-iteration runtime of 
a persistent collective, and $\tau_{\text{persistent}}$ the one-time cost of 
initialization and finalization. Persistence becomes advantageous once the number 
of iterations $N$ exceeds the break-even point:
\[
N_{\text{breakeven}} = 
\left\lceil 
\frac{\tau_{\text{persistent}}}{T_{\text{nonpersistent}} - T_{\text{persistent}}} 
\right\rceil.
\]
This equation shows that persistence pays off only after a sufficient number of 
repeated calls, when the one-time setup overhead is amortized by the savings in each subsequent iteration.

Together, these contributions demonstrate the potential of persistent RMA-based \texttt{Alltoallv} and provide valuable insights for building scalable communication libraries on modern HPC systems. Additionally, our discussion of trade-offs helps end users make informed decisions about when to adopt persistent RMA-based \texttt{Alltoallv} approaches in their group-oriented communications. All our approaches are available as open-source in a publicly accessible library maintained by the community; the repository link will be provided in the camera-ready version. %All our approaches are open source. and can be found in the MPI Advance library \cite{bienz2023mpiadvanceopensource}.

The rest of the paper is organized as follows. Section~\ref{Background} provides background on MPI Remote Memory Access (RMA) and related work. Section~\ref{Methodology} outlines the methodology and implementation details of the proposed persistent RMA-based \texttt{MPI\_Alltoallv} algorithms. Section~\ref{fence_init} describes the fence-based variant, Section~\ref{lock_persist} presents the lock-based RMA variants, and Section~\ref{alg:rma-fence-han} introduces the hierarchical fence-based variant. Section~\ref{benchmark} presents the first benchmark setup, while Section~\ref{Results} discusses performance results and analysis. Section~\ref{benchmark2} describes the second benchmark based on test suite sparse patterns. Finally, Section~\ref{Conclusion} concludes the paper and suggests directions for future work. %Acknowledgments are provided in Section~\ref{Acknowledgements}.

\section{Background}
\label{Background}
Bienz et al\hbox{.} \cite{bienz2023mpiadvanceopensource} propose MPI Advance, a suite of lightweight open-source libraries that improve message passing performance by 
building on existing MPI implementations. Their work introduces optimized collective and neighborhood collective operations, support for MPI-4 partitioned communication, and GPU-aware communication strategies. The MPI Advance library enables applications to benefit from advanced algorithms and new features without modifying the underlying MPI implementation, providing flexibility, portability, and access to performance improvements even on unmodified system MPIs.

Schuchart et al\hbox{.} \cite{schuchart2021quovadismpirma} discusses enhancements to the MPI RMA interface, with the aim of improving its efficiency for collective operations. The authors propose several modifications to MPI RMA, including the introduction of window duplication and memory handles to better meet the needs of large-scale parallel applications. These changes lead to significant performance improvements, particularly in terms of communication overlap and memory access optimization, which are crucial for minimizing latency in HPC environments.

Gerstenberger et al\hbox{.} \cite{Gerstenberger_2013} investigates the implementation of MPI-3 RMA, highlighting its potential to achieve superior performance compared to traditional methods. Their work emphasizes the advantages of one-sided communication, particularly in the context of large-scale parallel applications, where such methods reduce synchronization overhead and improve scalability. By utilizing MPI-3 RMA, the authors demonstrate significant enhancements in communication efficiency, leading to faster execution times and more efficient memory usage in distributed computing environments. 

Liu et al\hbox{.} \cite{liu2003rdma} propose a high-performance RDMA-based MPI implementation over InfiniBand that combines RDMA with traditional send/receive communication to improve scalability. Their design is effective across both small and large message sizes and supports efficient point-to-point as well as collective operations.

Tran et al\hbox{.} \cite{10833671} introduced OHIO, an optimized RDMA-based approach designed to enhance the scalability of \texttt{Alltoall} operations. By focusing on improving communication overlap, OHIO significantly reduces latency and increases throughput, which are crucial for achieving high performance in large-scale distributed systems. They attribute their performance improvements to cache thrashing that occurs inside network adapters. The study demonstrates that optimizing communication overlap can reduce bottlenecks commonly encountered in traditional \texttt{Alltoall} implementations, making OHIO a useful technique for improving the performance of collective communication operations in HPC environments.

Sur et al\hbox{.} \cite{10.1007/11602569_19} propose an RDMA-based design for the All-to-All broadcast operation, specifically targeting InfiniBand clusters. Their approach seeks to eliminate the overhead commonly associated with traditional communication methods, offering a more efficient solution for large-scale distributed systems. By leveraging RDMA,
their design significantly reduces latency, for larger message sizes. The authors demonstrate that their RDMA-based method outperforms conventional designs.

 Liu et al., Tran et al., and Sur et al., have demonstrated the benefits of RDMA for improving MPI communication performance, though with different emphases than our present work. Liu et al.\ integrate RDMA into the MPI implementation to accelerate point-to-point communication and collectives such as MPI\_Allgather, without addressing persistent collective interfaces or irregular Alltoallv patterns. Tran et al.\ improve the scalability of MPI\_Alltoall through optimized RDMA overlap and hierarchical techniques, but do not consider persistence or the amortization of initialization and synchronization costs across repeated iterations. Sur et al.\ propose an RDMA-based design for all-to-all broadcast on InfiniBand, but similarly do not address persistent RMA collectives or irregular communication patterns. In contrast, our work specifically targets persistent MPI RMA based Alltoallv designs, separating one-time setup from per-iteration execution to amortize overheads.

Namugwanya et al.\@ \cite{namugwanya2023collectiveoptimizedffts} explore novel implementations of the \texttt{MPI\_Alltoallv} operation to improve the performance and scalability of FFT solvers, such as HeFFTe~\cite{heffte}. This study, investigates the impact of optimized communication patterns in parallel computing applications, particularly within HPC environments. The authors demonstrate significant improvements in communication efficiency. 

Bienz et al\hbox{.}~\cite{ModelingDataMovement} investigate various contributors to data movement costs in parallel systems, highlighting factors such as system architecture, job partitioning, and task adjacency. Their study emphasizes the role of accurate performance models in identifying communication bottlenecks. Because of the complexities introduced by modern heterogeneous architectures where multiple pathways exist for inter-GPU communication, they proposed models to evaluate different inter-node communication strategies. Notably, they analyzed the tradeoffs between GPU-Direct and CPU-mediated communication, and introduced a novel optimization approach that utilizes all CPU cores per node to enhance internode communication. %Their findings demonstrate significant performance gains for MPI collective operations.

Jocksch et al\hbox{.}~\cite{Optimized} propose an optimized \texttt{Alltoall} communication algorithm designed for multicore systems, specifically to improve FFT performance using the pencil decomposition. The approach takes advantage of the hybrid parallelism present in modern supercomputers, which combine shared and distributed memory models. Their work advocates for enhancements to the MPI standard to support such communication patterns and showcases promising improvements.

%Kumar et al\hbox{.}~\cite{Amr_Scaling_Alltoall_Collective_on_Multi-core_Systems} focused on enhancing the scalability of \texttt{Alltoall} collectives on multi-core systems. They introduce a design that avoids reliance on a leader process, aiming instead to fully utilize all available cores across nodes for faster inter-node communication. In addition, they analyzed how different network interface card (NIC) architectures impact the performance of MPI \texttt{Alltoall} operations.

Kumar et al\hbox{.}~\cite{Amr_Scaling_Alltoall_Collective_on_Multi-core_Systems} focused on designing MPI\_Alltoall schemes tailored to multi-core architectures. They show that no single communication strategy performs optimally across all architectures, motivating architecture-aware designs. For example, onloaded implementations can exploit multiple cores to improve network utilization, while offloaded interfaces can use aggregation to reduce congestion on multi-core systems. Their approach employs shared-memory aggregation techniques and achieves up to a 55\% reduction in MPI\_Alltoall time.% for 512-byte messages, as well as a 33\% speedup for the CPMD application.

Hofmann and Runger~\cite{Amr_An_InPlaceAlgorithmForIrregularAllToAllCommunicationwithLimitedMemory} developed a novel in-place \texttt{MPI\_Alltoallv} algorithm that overcomes the traditional constraint of uniform message sizes. Their approach allows for varying counts and displacements of messages while still using shared in-place buffers. This flexibility supports irregular data exchanges, making the algorithm suitable for memory-constrained applications that require dynamic communication patterns.

Mamadou et al\hbox{.}~\cite{Amr_ADynamicSolutionforEfficientMPICollectiveCommunications} presented DYN \texttt{Alltoall}, a dynamic implementation of \texttt{MPI\_Alltoall} based on performance predictions using the P-LogP model. Unlike static methods, DYN \texttt{Alltoall} adapts to system and network variability by selecting the most suitable communication algorithm at runtime. This adaptability enhances the robustness and efficiency of collective operations in varying execution environments.

Besta et al\hbox{.}~\cite{besta2020acceleratingirregularcomputationshardware} propose Atomic Active Messages (AAM), a communication mechanism designed to accelerate irregular graph computations across both shared and distributed memory architectures. The central idea behind AAM is the use of hardware transactional memory (HTM) to simplify and speed up processing of irregular data structures in parallel environments. They demonstrated how techniques such as coarsening and coalescing enable efficient execution of hardware transactions, leading to significant performance gains. Their evaluation on Intel Haswell and IBM Blue Gene/Q systems highlighted trade-offs in HTM configurations and showed how AAM can be applied to improve the performance of existing graph processing frameworks.

Dosanjh et al\hbox{.}\cite{Dosanjh} introduce RMA-MT, the first publicly available suite of proxy applications and microbenchmarks designed to evaluate multi-threaded MPI RMA performance. Their study systematically examines how different MPI implementations behave under thread-multiple RMA execution and demonstrates how the benchmark suite can be used to assess both performance and correctness.%, as well as to guide MPI implementation development.

White\cite{10.1145/3731599.3767389} evaluates GPU-aware all-to-all communication at extreme scale using a combination of persistent collectives
(\texttt{MPI\_Alltoall\_init}), one-sided communication (\texttt{MPI\_Get}), and hierarchical node-aware algorithms. While the study demonstrates that hierarchical and RMA-inspired designs can significantly reduce latency for large messages, persistence is not the primary optimization mechanism, and the proposed approaches do not provide a persistent RMA collective interface or address irregular \texttt{MPI\_Alltoallv} patterns.

Persistent communication has also been recognized as a key mechanism for improving 
the efficiency of RMA-based collectives. Morgan et al.~\cite{Morgan_Persistent} 
introduced a systematic exploration of persistent collective operations in MPI, with a 
particular emphasis on how planning and setup costs can be distributed across multiple 
iterations. Their work demonstrated that persistent initialization of collectives 
eliminates the need to repeatedly allocate windows, compute displacements, and exchange 
metadata on every invocation. Instead, these tasks are performed once and stored in 
reusable request objects, enabling subsequent iterations to invoke only lightweight 
start and wait operations. This design substantially reduces per-iteration overhead and 
improves scalability for iterative workloads such as stencil computations and solvers. 
The insights from this study directly relate to the recent developments of this paper in persistent 
RMA-based \texttt{Alltoallv} algorithms, where separation of initialization from execution is beneficial for scalability purposes.

\section{Methodology}
\label{Methodology}
To evaluate the effectiveness of Remote Memory Access (RMA) in collective communication, we
implemented and analyzed multiple RMA-based alternatives of the \texttt{MPI\_Alltoallv} routine.
Our methodology focuses mainly on two synchronization approaches: fence-based and lock-based RMA
implementations. Each approach is designed to leverage MPI’s one-sided (\texttt{MPI\_Put}) communication features
while addressing the scalability challenges of irregular data exchange. \textbf{All our
RMA variants are developed as \emph{persistent collectives}}: initialization (\textsc{init})
performs a one-time setup that creates or reuses the receive window, exchanges remote
displacements, computes datatype sizes, and stores all metadata in a persistent request;
execution then proceeds via \textsc{start}/\textsc{wait}  epochs; finalization (\textsc{finalize, free}) releases resources. When the
total receive bytes remain constant, the same window and request are reused across iterations,
avoiding repeated metadata exchange and window creation; if sizes change, the window is recreated. This separation lets us measure overheads and apply a break-even analysis to determine when persistence pays off and how much time is saved. 

In the following subsections, we detail the core algorithms (Fence persistent and Lock persistent), as well as the Fence\_hierarchy\_persistent variant, which follows the same synchronization semantics as the fence-based algorithm but reorders remote and local transfers to enable overlap. Because this strategy behaves identically for fence- and lock-based synchronization, we present only the fence-hierarchy-aware variant for clarity.

\subsection{Fence-based Persistent RMA Alltoallv}
\label{fence_init}
As discussed earlier, to support efficient reuse of communication resources across multiple iterations, we developed an initialization function, for persistent RMA-based \texttt{Alltoallv} operations. Unlike traditional implementations that allocate and configure communication resources on every invocation, this function performs a one-time setup of all metadata and RMA resources, storing them in a reusable \texttt{MPIX\_Request} object.

The routine queries the sizes of the send and receive datatypes, converts all counts and displacements into byte units, and computes the total receive buffer size. If the currently cached RMA window does not match the required size, the existing window is freed and a new window is created over \texttt{recvbuf}. It then performs an \texttt{MPI\_Alltoall} on \texttt{rdispls} to determine the remote byte offsets (\texttt{put\_displs}) within each target process’ exposed window. These offsets, along with the computed send and receive sizes and communication metadata, are stored in the request object as summarized Algorithm~\ref{alg:rma-fence}.

The persistent design enables subsequent calls to invoke only \texttt{rma\_start} and \texttt{rma\_wait} without repeating setup.
\texttt{rma\_start} opens the epoch using \texttt{MPI\_Win\_fence} and performs the data transfer with \texttt{MPI\_Put}, while \texttt{rma\_wait} closes the epoch with a second \texttt{MPI\_Win\_fence}. This separation of initialization and execution phases reduces overhead and improves efficiency, especially in applications with repeated communication phases, such as stencil computations or iterative solvers.

Algorithm ~\ref{alg:rma-fence-han} presents the Fence hierarchy persistent algorithm, which extends the fence-based RMA approach by distinguishing between remote and intra-node targets and reordering data transfers to enable overlap while preserving the same fence synchronization semantics.

\begin{comment}
\begin{algorithm*}
    
\caption{Fence-Based Persistent \texttt{Alltoallv}: \textsc{init}/\textsc{start}/\textsc{wait}}
\label{alg:rma-fence}
\begin{verbatim}

INIT:
  Enter meta data via function ALLTOALLV_RMA_FENCE_INIT(sendbuf, sendcounts, 
  
    sdispls,sendtype,recvbuf, recvcounts, rdispls, recvtype,xcomm, xinfo, request_ptr)
  
  1. Query rank, size; allocate request structure
  2. Compute total_recv_bytes ← sum_i recvcounts[i] × sizeof(recvtype), for i = 0 ... P-1
  
  3. MPI_WIN_CREATE(recvbuf, total_recv_bytes, 1, MPI_INFO_NULL,
                       xcomm.global_comm, &xcomm.win)
  4. Note: Implementation may cache/reuse the window when total_recv_bytes is unchanged.
    
  5. MPI_Alltoall(rdispls, 1, MPI_INT, put_displs, 1, MPI_INT, comm)
  6. Convert count/displs to bytes.
  7. MPI_Alltoall(sendcounts) to get incoming_send_counts_from[sending processi]) 
  8. Ensure for each incoming data from sender i to  P :recvcounts[P] >= sendcounts_from_i[P] 
  9. Store metadata {win, displs, sizes, types, comm} in request object (ptr)
  10. Return
     

START:
  11. MPI_Win_fence(MPI_MODE_NOSTORE | MPI_MODE_NOPRECEDE, win)
  12. For each process p:
       MPI_Put(sendbuf + sdispls[p], sendcounts[p], MPI_BYTE,
               p, put_displs[p], sendcounts[p], MPI_BYTE, win)
 13. Return            

WAIT:
 14. MPI_Win_fence(MPI_MODE_NOPUT | MPI_MODE_NOSUCCEED, win)
 15. Return
  
FREE:
 16.MPIX_COMM_WIN_FREE(win) // old window invalid
 17.Return
  
\end{verbatim}
\end{algorithm*}
\par\noindent\ignorespaces

\end{comment}

\begin{algorithm}
\caption{Fence-Based Persistent \texttt{Alltoallv}: \textsc{init}/\textsc{start}/\textsc{wait}}
\label{alg:rma-fence}
\footnotesize
\begin{algorithmic}[1]

\Statex \textbf{INIT:}
\Statex Enter metadata via function:
\Statex \begin{minipage}[t]{\linewidth}
\texttt{ALLTOALLV\_RMA\_FENCE\_INIT(}\\
\hspace*{\algorithmicindent}\texttt{sendbuf, sendcounts, sdispls, sendtype,}\\
\hspace*{\algorithmicindent}\texttt{recvbuf, recvcounts, rdispls, recvtype,}\\
\hspace*{\algorithmicindent}\texttt{xcomm, xinfo, request\_ptr)}
\end{minipage}

\State Query rank, size; allocate request structure
\State Compute \texttt{total\_recv\_bytes} $\leftarrow \sum_{i=0}^{P-1} \texttt{recvcounts}[i]\times \texttt{sizeof(recvtype)}$

\Statex \begin{minipage}[t]{\linewidth}
\texttt{MPI\_WIN\_CREATE(}\\
\hspace*{\algorithmicindent}\texttt{recvbuf, total\_recv\_bytes, 1, MPI\_INFO\_NULL,}\\
\hspace*{\algorithmicindent}\texttt{xcomm.global\_comm, \&xcomm.win)}
\end{minipage}

\State Note: Implementation may cache/reuse the window when \texttt{total\_recv\_bytes} is unchanged
\State \texttt{MPI\_Alltoall(rdispls, 1, MPI\_INT, put\_displs, 1, MPI\_INT, comm)}
\State Convert counts/displs to bytes
\State \texttt{MPI\_Alltoall(sendcounts)} to get \texttt{incoming\_send\_counts\_from[sender\_i]}
%\State Ensure for each incoming data from sender $i$ to $p$: \texttt{recvcounts[p] $\ge$ sendcounts\_from\_i[p]}
\State Store metadata \{\texttt{win, displs, sizes, types, comm}\} in request object (\texttt{ptr})
\State Return

\Statex
\Statex \textbf{START:}
\State \texttt{MPI\_Win\_fence(MPI\_MODE\_NOSTORE $|$ MPI\_MODE\_NOPRECEDE, win)}
\For{each process $p$}
  \Statex \begin{minipage}[t]{\linewidth}
  \texttt{MPI\_Put(}\\
  \hspace*{\algorithmicindent}\texttt{sendbuf + sdispls[p], sendcounts[p], MPI\_BYTE,}\\
  \hspace*{\algorithmicindent}\texttt{p, put\_displs[p], sendcounts[p], MPI\_BYTE, win)}
  \end{minipage}
\EndFor
\State Return

\Statex
\Statex \textbf{WAIT:}
\State \texttt{MPI\_Win\_fence(MPI\_MODE\_NOPUT $|$ MPI\_MODE\_NOSUCCEED, win)}
\State Return

\Statex
\Statex \textbf{FREE:}
\State \texttt{MPIX\_COMM\_WIN\_FREE(win)} %\Comment{old window invalid}
\State Return

\end{algorithmic}
\end{algorithm}

\begin{algorithm}
\caption{Fence-Hierarchy Persistent \texttt{Alltoallv}: \textsc{init}/\textsc{start}/\textsc{wait}}
\label{alg:rma-fence-han}
\footnotesize
\begin{algorithmic}[1]

\Statex \textbf{INIT:}
\Statex Enter metadata via function:
\Statex \begin{minipage}[t]{\linewidth}
\texttt{ALLTOALLV\_RMA\_FENCE\_INIT\_HIERARCHY(...)}
\end{minipage}
\State Query rank, size; allocate request structure
\State Compute \texttt{total\_recv\_bytes} = $\sum_i$ \texttt{recvcounts[i]} * \texttt{sizeof(recvtype)}, for $i = 0 \ldots P-1$
\Statex \begin{minipage}[t]{\linewidth}
\texttt{MPI\_Win\_create(recvbuf, total\_recv\_bytes, 1, MPI\_INFO\_NULL,}\\
\hspace*{\algorithmicindent}\texttt{xcomm.global\_comm, \&xcomm.win)}
\end{minipage}
\State \texttt{MPI\_Alltoall(rdispls, 1, MPI\_INT, put\_displs, 1, MPI\_INT, comm)}
\State Convert displs/sizes to byte units
\State \texttt{MPI\_Alltoall(sendcounts)} to get \texttt{incoming\_send\_counts\_from[sending process])}
\State Ensure for each incoming data from sender $i$ to $P$ : \texttt{recvcounts[P]} $\ge$ \texttt{sendcounts\_from\_i[P]}
\State Split communicator by node (\texttt{MPI\_COMM\_TYPE\_SHARED})
\Statex \hspace*{\algorithmicindent}$\rightarrow$ identify which ranks are local vs. remote
\State Build target lists:
\Statex \hspace*{\algorithmicindent}\texttt{remote\_targets = ranks on other nodes}
\Statex \hspace*{\algorithmicindent}\texttt{local\_targets  = ranks on same node}
\State Store \{\texttt{win, displs, sizes, local/remote lists}\} in request
\State Return

\Statex
\Statex \textbf{START:}
\State \texttt{MPI\_Win\_fence(MPI\_MODE\_NOPRECEDE, win)}
\Statex For each $i$ in \texttt{remote\_targets:}
\Statex \begin{minipage}[t]{\linewidth}
\hspace*{\algorithmicindent}\texttt{MPI\_Put(sendbuf + sdispls[p], send\_counts[p], MPI\_BYTE,}\\
\hspace*{2\algorithmicindent}\texttt{i, put\_displs[p],send\_counts[p], MPI\_BYTE, win)}
\end{minipage}
\Statex For each $i$ in \texttt{local\_targets:}
\Statex \begin{minipage}[t]{\linewidth}
\hspace*{\algorithmicindent}\texttt{MPI\_Put(sendbuf + sdispls[p], send\_counts[p], MPI\_BYTE,}\\
\hspace*{2\algorithmicindent}\texttt{i, put\_displs[p], send\_counts[p], MPI\_BYTE, win)}
\end{minipage}
\State Return

\Statex
\Statex \textbf{WAIT:}
\State \texttt{MPI\_Win\_fence(MPI\_MODE\_NOPUT $|$ MPI\_MODE\_NOSUCCEED, win)}
\State Return

\Statex
\Statex \textbf{FREE:}
\State \texttt{MPIX\_COMM\_WIN\_FREE(win)}
\State Return

\end{algorithmic}
\end{algorithm}

\subsection{Lock-based Persistent RMA Alltoallv}
\label{lock_persist}
Similar to our previous fence variant explanation, to support repeated use of passive-target synchronization in persistent RMA-based \texttt{Alltoallv} communication, we implemented a persistent setup routine named \texttt{alltoallv\_rma\_lock\_init}. This function corresponds to \textbf{Algorithm} ~\ref{alg:lock-init} and performs a one-time setup of all communication metadata. These resources are stored in a reusable \texttt{MPIX\_Request} object, allowing communication to be triggered efficiently in later iterations. These resources include the RMA window handle, byte-converted send and receive sizes, and remote byte-displacement metadata; by caching this state in a reusable \texttt{MPIX\_Request}, we avoid repeated window setup and metadata exchange on each iteration, which reduces per-iteration runtime as detailed in the Section \ref{Results}.

Unlike the fence-based version, which relies on collective synchronization via \texttt{MPI\_Win\_fence}, the lock-based variant uses exclusive and shared locking to manage memory exposure. Specifically, the initialization function begins with an exclusive self-lock (\texttt{MPI\_Win\_lock}) to ensure that the receive buffer can be safely prepared and the window can be created. If the RMA window is uninitialized or no longer matches the receive buffer size, it is freed and recreated with \texttt{MPI\_Win\_create} using the appropriate new size and displacement unit. The new window handle is then stored in \texttt{xcomm}.

Following this, the function initializes displacement arrays. An \texttt{MPI\_Alltoall} exchange is performed to collect the \texttt{rdispls} from all peers and compute the corresponding byte-level remote offsets (\texttt{put\_displs}). These offsets are later used to target the correct position in each remote process’s receive buffer.

The actual data transfer is handled by the function \texttt{rma\_lock\_start}(Start) (\textbf{Algorithm \ref{alg:lock-init}}), which releases the exclusive self-lock and initiates a shared lock-all epoch using \texttt{MPI\_Win\_lock\_all}. It then posts a series of \texttt{MPI\_put} operations to each peer where \texttt{send\_sizes[i] > 0}.

The \texttt{rma\_lock\_wait} (Wait) function finalizes the communication epoch by calling \texttt{MPI\_Win\_unlock\_all}. Then we ensure that every process has finished their access epochs with the \texttt{MPI\_Barrier}, then calls and reacquires an exclusive lock. This step ensures that the receive buffer is in a consistent state for reading.

Finally, during the FREE phase, each process first releases its exclusive self-lock on the window, then frees the persistent request object, and finally deallocates the RMA window using \texttt{MPIX\_COMM\_WIN\_FREE}. This sequence ensures that all access epochs are closed before resources are released.

\begin{algorithm}
\caption{Lock-Based Persistent \texttt{Alltoallv}: \textsc{init}/\textsc{start}/\textsc{wait}}
\label{alg:lock-init}
\footnotesize
\begin{algorithmic}[1]

\Statex \textbf{INIT:}
\Statex Enter metadata via function:
\Statex \begin{minipage}[t]{\linewidth}
\texttt{ALLTOALLV\_RMA\_LOCK\_INIT(}\\
\hspace*{\algorithmicindent}\texttt{sendbuf, sendcounts, sdispls, sendtype,}\\
\hspace*{\algorithmicindent}\texttt{recvbuf, recvcounts, rdispls, recvtype,}\\
\hspace*{\algorithmicindent}\texttt{xcomm, xinfo, request\_ptr)}
\end{minipage}

\State Query rank, size; allocate request structure
\State Compute \texttt{total\_recv\_bytes} $\leftarrow \sum_{i=0}^{P-1} \texttt{recvcounts}[i]\times \texttt{sizeof(recvtype)}$

\Statex \begin{minipage}[t]{\linewidth}
\texttt{MPI\_WIN\_CREATE(}\\
\hspace*{\algorithmicindent}\texttt{recvbuf, total\_recv\_bytes, 1, MPI\_INFO\_NULL,}\\
\hspace*{\algorithmicindent}\texttt{xcomm.global\_comm, \&xcomm.win)}
\end{minipage}

\State Note: Implementation may cache/reuse the window when \texttt{total\_recv\_bytes} is unchanged

\State \texttt{MPI\_Win\_lock(MPI\_LOCK\_EXCLUSIVE, rank, 0, xcomm.win)} \Comment prevent window access while metadata is configured

\Statex \begin{minipage}[t]{\linewidth}
\texttt{MPI\_Alltoall(rdispls, 1, MPI\_INT, put\_displs, 1, MPI\_INT, xcomm.global\_comm)}
\end{minipage}

\State Convert count/displs to bytes

\Statex \begin{minipage}[t]{\linewidth}
\texttt{MPI\_Alltoall(sendcounts\_bytes)} to get \\
\hspace*{\algorithmicindent}\texttt{incoming\_send\_bytes\_from[sending process]}
\end{minipage}

\State Ensure for each incoming data from sender $i$ to $P$: \texttt{recvcounts[P] $\ge$ sendcounts\_from\_i[P]}
\State Store metadata \{\texttt{xcomm.win, displs, sizes, types, xcomm}\} in \texttt{request\_ptr}

\Statex
\Statex \textbf{START (RMA\_LOCK\_START):}
\State \texttt{MPI\_Win\_unlock(rank, xcomm.win)} \Comment{release self lock from init}
\State \texttt{MPI\_Win\_lock\_all(0, xcomm.win)} \Comment{begin access epoch}

\For{each process $p$}
  \Statex \begin{minipage}[t]{\linewidth}
  \texttt{MPI\_Put(}\\
  \hspace*{\algorithmicindent}\texttt{sendbuf + sdispls[p], sendcounts[p], MPI\_BYTE,}\\
  \hspace*{\algorithmicindent}\texttt{p, put\_displs[p], sendcounts[p], MPI\_BYTE,}\\
  \hspace*{\algorithmicindent}\texttt{xcomm.win)}
  \end{minipage}
\EndFor
\State Return

\Statex
\Statex \textbf{WAIT (RMA\_LOCK\_WAIT):}
\State \texttt{MPI\_Win\_unlock\_all(xcomm.win)} \Comment{end epoch, ensure remote completion}
\State \texttt{MPI\_Win\_lock(MPI\_LOCK\_EXCLUSIVE, rank, 0, xcomm.win)}
\State Return

\Statex
\Statex \textbf{FREE:}
\State \texttt{MPI\_Win\_unlock(rank, xcomm.win)}
\State \texttt{MPIX\_COMM\_WIN\_FREE(xcomm)}
\State Return

\end{algorithmic}
\end{algorithm}

\section{Benchmark 1 Description}
\label{benchmark}
To evaluate the performance and correctness of the persistent RMA-based \texttt{Alltoallv} implementations, we developed a benchmark built on the \texttt{MPI-Advance} interface. The benchmark supports both strong and weak scaling.The benchmark uses uniform counts and displacements across ranks to keep the send and receive buffer sizes constant across iterations, enabling reuse of the same RMA window and associated communication metadata. This allows the persistent RMA paths to reuse a single initialized request object across iterations. By isolating one time setup costs (window creation and metadata preparation) from per-iteration execution, the persistent design amortizes initialization overhead across repeated invocations.
 
Each process initializes its send buffer such that all elements destined for a given rank $i$ are set to the sender's rank ID. This predictable pattern allows element-wise validation on the receiving side.

The benchmark measures average runtime over 1000 iterations for each implementation:
\begin{itemize}
    \item \texttt{MPI\_Alltoallv} (baseline)
    \item \texttt{alltoallv\_rma\_winfence\_init} (persistent fence-based initialization)
    \item \texttt{alltoallv\_rma\_lock\_init} (persistent lock-based initialization)
    \item \texttt{MPIX\_Start()}(starts actual persistent calls)
    \item \texttt{MPIX\_Wait()}(Waits on actual persistent calls)
\end{itemize}

Timing is performed using \texttt{MPI\_Wtime()}, with  maximum time reported via \texttt{MPI\_Reduce()} using \texttt{MPI\_MAX} to capture worst-case communication time. We insert \texttt{MPI\_Barrier()}
before each timed region.

In both RMA variants, the initialization function is first invoked in a loop to simulate the startup overhead, followed by persistent communication using \texttt{MPIX\_Start()} and \texttt{MPIX\_Wait()}. For correctness, the receive buffers are validated element-by-element after each variant completes. The benchmark reports any mismatches.
%This is explained earlier
%For both variants, the  window is reused across iterations whenever the total receive
%bytes are unchanged; if the size changes, the old window is freed and recreated.

\section{Results}
\label{Results}

All experiments were conducted on the Dane supercomputer of LLNL, which is equipped with an InfiniBand interconnect.

\begin{figure}
    \centering
    \includegraphics[width=0.98\columnwidth]{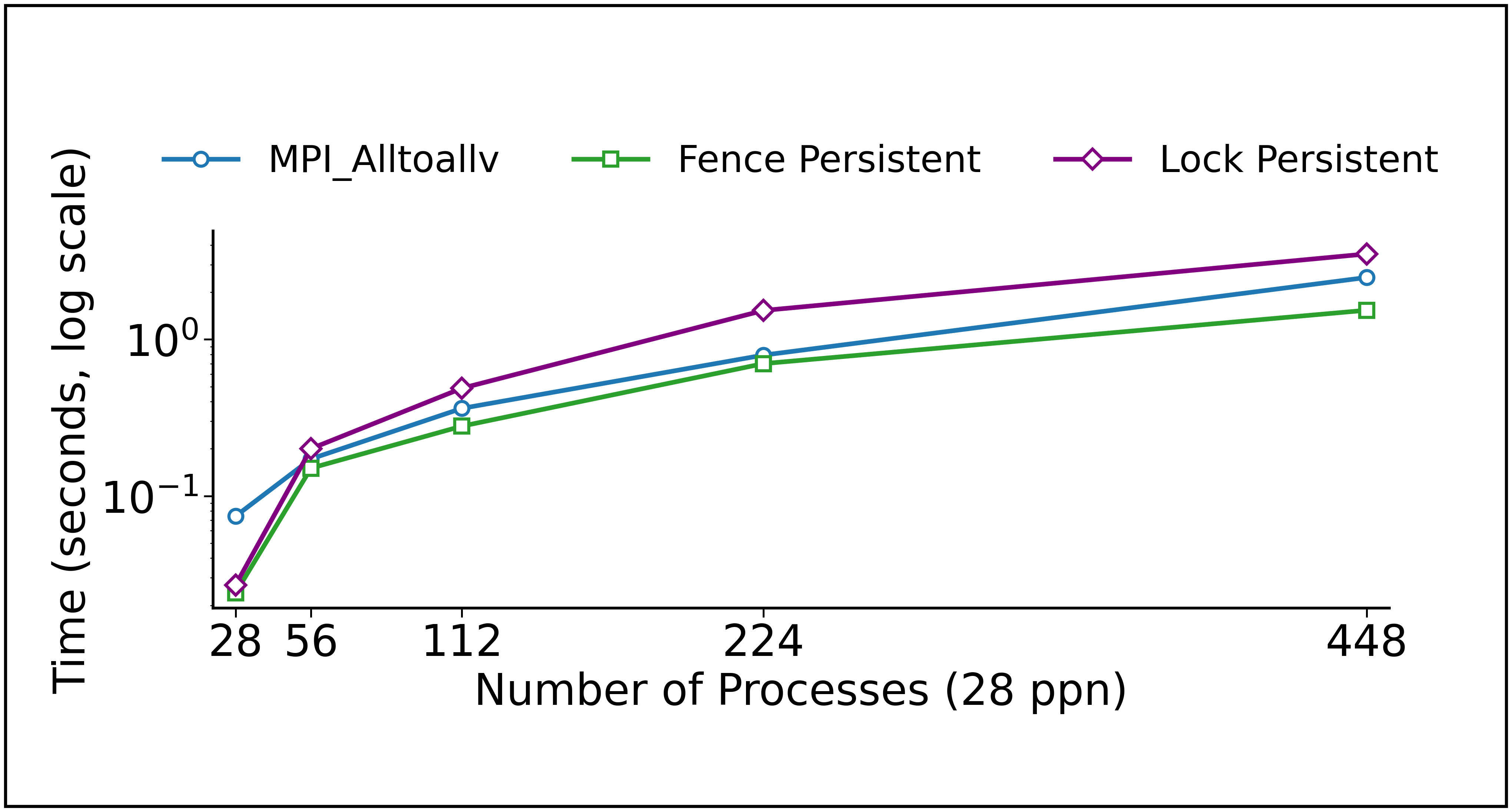}
    \caption{ 2,097,152 bytes per process (weak scaling):Performance comparison of standard \texttt{MPI\_Alltoallv} against our proposed persistent RMA-based algorithms (Fence and Lock persistent) across varying process counts on the Dane supercomputer.}
    \label{fig:rma_persistent_scaling}
\end{figure}

Figure~\ref{fig:rma_persistent_scaling} shows the \textbf{weak scaling} behavior of three 
\texttt{Alltoallv} implementations: the baseline (\texttt{MPI\_Alltoallv} (MVAPICH2)), 
the Fence persistent RMA algorithm, and the Lock persistent RMA algorithm. 

At low process counts (28), the Fence Persistent algorithm achieves the best performance, 
completing communication in 24--27~ms compared to 74~ms for the baseline. 
This advantage highlights the benefit of reduced synchronization costs when only a small number processes are doing fence operations. 
The Lock Persistent algorithm performs comparably at 56 processes but begins to perform differently as number of process increases. 

As the number of processes grows (112 to 224), execution times increase for all implementations due to higher communication overhead. 
Fence persistent maintains its advantage, scaling more efficiently than both the baseline 
and Lock persistent. 
At 224 processes, Fence persistent completes in 700~ms compared to 794~ms for the baseline, 
while Lock persistent requires 1.53~s. 

At the largest scale tested (448 processes), the performance gap is highest. 
Fence Persistent sustains efficient scaling with a runtime of 1.54~s, whereas the baseline requires 
2.49~s and Lock persistent incurs 3.52~s. 
Thus, Fence persistent is 0.95~s faster than the baseline, representing an improvement of approximately 38\%. 
%got 38% by computing the relative improvement over the baseline runtime(2.49−1.54=0.95, (0.9(improvement)/baseline(2.49))*100=38%.
The performance of the Lock-based algorithm is because of its need for multiple synchronization steps: 
each target requires a lock/unlock pair, along with additional synchronization to guarantee memory consistency before buffers can be accessed by the application. 
In contrast, the fence-based approach requires only a single initial fence and a final fence per communication epoch, reducing synchronization overhead and enabling better scalability.

\begin{figure}
    \centering
    \includegraphics[width=1.00\columnwidth]{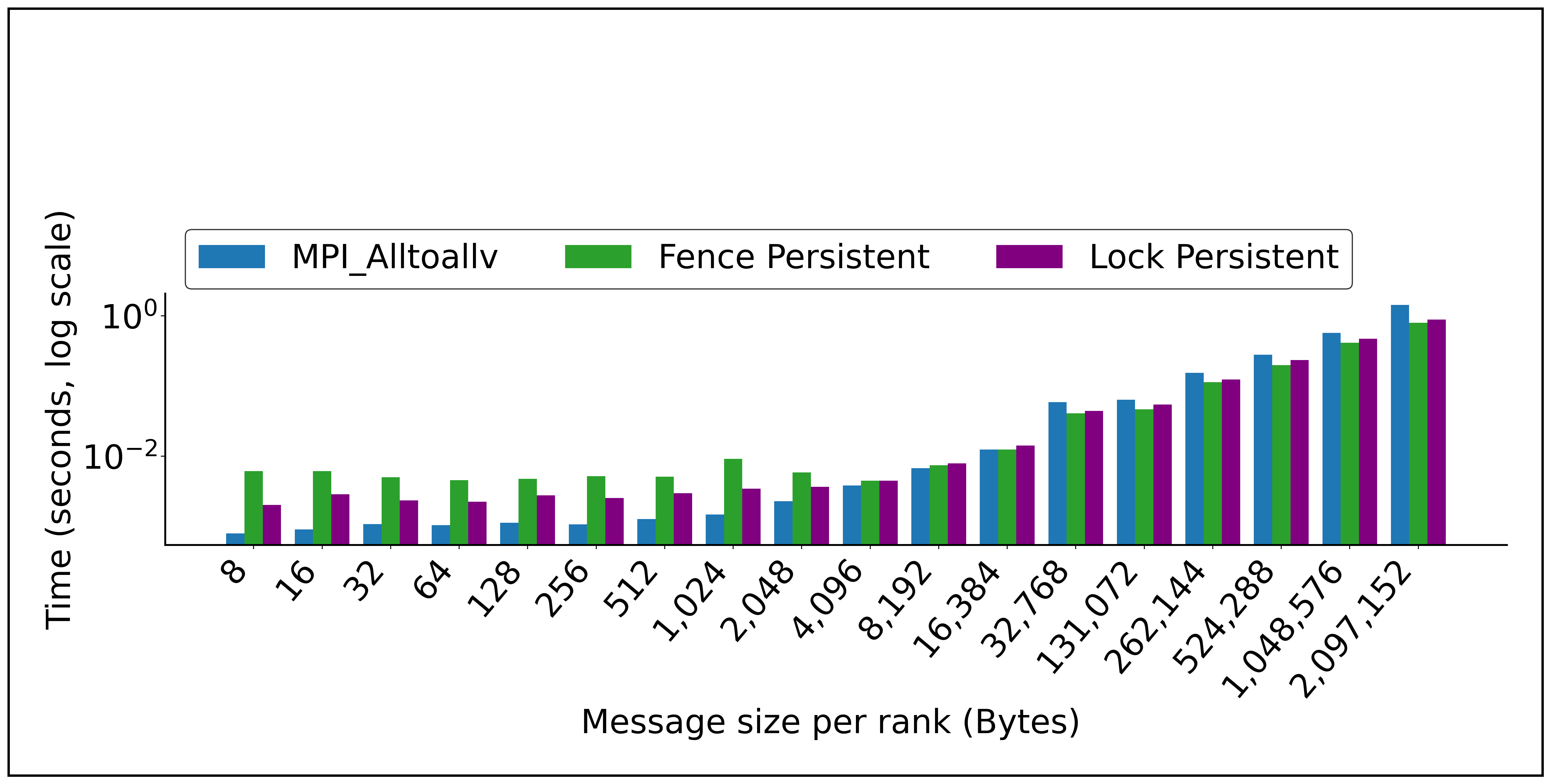}
    \caption{ Varying Message size per process (bytes) on 8 nodes, 224 procs, 28 ppn.}
    \label{newbar_PMPI_vs_FencePersist_vs_LockPersist}
\end{figure}

\textbf{Figure~\ref{newbar_PMPI_vs_FencePersist_vs_LockPersist}} compares  runtimes  of the
baseline MPI \texttt{Alltoallv}, a Fence persistent RMA Alltoallv, and a Lock persistent RMA Alltoallv across varrying message sizes on 8 nodes (224 processes, 28~ppn).
At smaller message sizes (less or equal to 8,192 bytes) Native \texttt{MPI\_Alltoallv} performs better than other algorithms.
As message size increase from (32,768 bytes beyond), the fence-persistent variant begins to  outperform the baseline by
reusing initialization, setup data.

As prior discussed we model the cost of a persistent  as
\begin{equation}
T_{\text{persist,total}} = T_{\text{init}} + N \cdot T_{\text{persist}},
\end{equation}
where \(N\) denotes the number of repeated iterations that reuse the same communication metadata, 
and \(T_{\text{init}}\) represents the one-time cost of initializing and finalizing the persistent request. 

The non-persistent baseline incurs
\begin{equation}
T_{\text{base,total}} = N \cdot T_{\text{MPI}},
\end{equation}
where \(T_{\text{MPI}}\) is the runtime of a standard \texttt{MPI\_Alltoallv} call.

Persistence becomes advantageous when \(T_{\text{persist,total}} < T_{\text{base,total}}\), 
yielding the break-even point
\begin{equation}
N_{\text{breakeven}}
 = \left\lceil
   \frac{T_{\text{init}}}{T_{\text{MPI}} - T_{\text{persist}}}
   \right\rceil ,
\end{equation}
where \(T_{\text{persist}}\) refers to the \emph{start+wait} runtime after initialization.

Applying this model to the measured runtimes in Figure ~\ref{newbar_PMPI_vs_FencePersist_vs_LockPersist} shows that for all message sizes 
\(\ge 32{,}768\)~bytes, \(\Delta T = T_{\text{MPI}} - T_{\text{persist}}\) is positive for our persistent variants, 
and the computed \(N_{\text{breakeven}} = 1\). 
This indicates that the one-time initialization overhead is recovered within the first invocation, 
providing immediate benefit.

At 32{,}768~bytes, Fence persistent reduces runtime from 0.0588\,s to 0.0410\,s, saving 
0.0178\,s (30.2\%) per call, while Lock persistent saves 0.0148\,s (25.1\%). 
Savings remain consistent at 131{,}072~bytes with 27.1\% improvement for Fence and 14.7\% for Lock. 
For larger messages the advantage grows: at 1{,}048{,}576~bytes, Fence persistent saves 0.158\,s (27.6\%) 
and Lock 0.100\,s (17.5\%) per iteration, and at 2{,}097{,}152~bytes the reductions reach 
0.637\,s (44.3\%) and 0.550\,s (38.3\%), respectively.

These improvements arise from \textbf{cached metadata} established during 
\texttt{init}, displacements, datatype decoding, 
and RMA descriptors are reused across invocations, eliminating repeated setup.

\subsection{Test Suite Sparse Benchmark (Benchmark 2)}
\label{benchmark2}
To evaluate persistent RMA-based \texttt{MPI\_Alltoallv} under irregular communication we use Suite Sparse Benchmark. The Benchmark uses sparse matrices from the SuiteSparse collection to generate various communication patterns based on matrix sparsity. Each matrix creates an irregular \texttt{MPI\_Alltoallv} exchange. Processes pack their send buffers in rank order using a predictable data pattern, enabling direct element-wise validation of received data against the baseline \texttt{MPI\_Alltoallv} (MVAPICH2). We also evaluated our implementations using Open MPI 4; however, the observed performance trends and relative ordering of algorithms were qualitatively similar to those obtained with MVAPICH2. For clarity, we therefore present results only for MVAPICH2 in this paper. Timings are collected using \texttt{MPI\_Wtime()} after a warm-up phase, with maximum time reported via \texttt{MPI\_Reduce} to capture worst-case. When receive sizes remain constant, RMA windows and persistent requests are reused across iterations, reflecting iterative application workloads like in FFTs.

%\subsection{variation of the algorithms on suite sparse pattern}

\begin{figure}
    \centering
    \includegraphics[width=1.00\columnwidth]{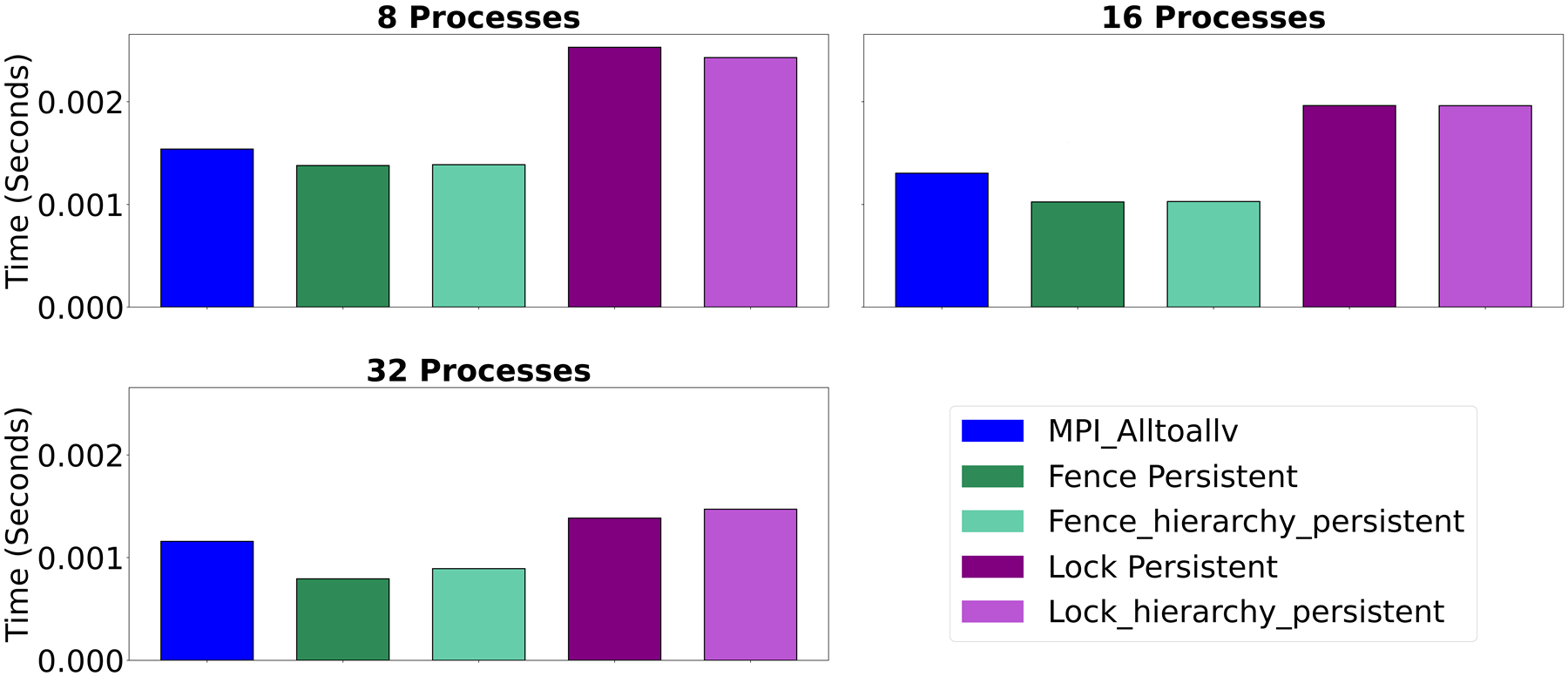}
    \caption{hugetrace-00020 (47,997,626 nnz), performance comparison of MPI\_Alltoallv and persistent RMA methods}
    \label{hugetrace_00020_1x3_grid_with_nnz.png}
\end{figure}
 
\begin{figure}
    \centering
    \includegraphics[width=1.00\columnwidth]{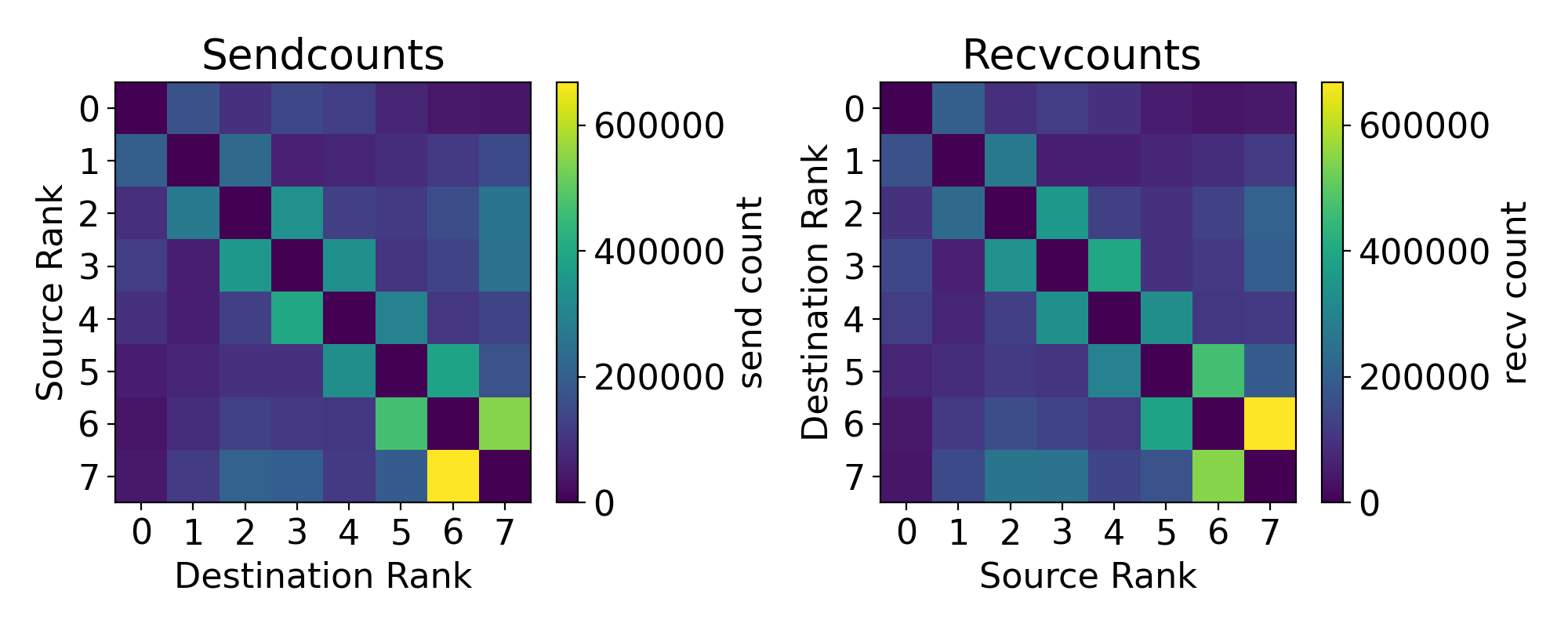}
    \caption{ 8 Nodes 8 Processes, analyzing the suite sparse pattern for the hugetrace-00020 matrix}
    \label{hugetrace-00020}
\end{figure}
Figure~\ref{hugetrace_00020_1x3_grid_with_nnz.png}, shows that the two \texttt{winfence}-based persistent variants 
(Fence persistent and Fence\_hierarchy\_persistent) have similar 
performance because they share the same global synchronization mechanism; their primary 
difference is the ordering of \texttt{MPI\_Put} operations as seen in algorithm ~\ref{alg:rma-fence-han}.

In contrast, the lock-based persistent methods perform worse due to the characteristics of passive target synchronization. Passive target RMA requires each origin process to acquire a per-target lock before issuing \texttt{MPI\_Put} operations. When many processes access the same target, these locks introduce serialization and additional synchronization overhead.

This effect is amplified for large or imbalanced messages: processes transferring more data hold the window lock for a longer period because the epoch cannot be released until all \texttt{MPI\_Put} operations complete. As RDMA transfer time grows with message size, other peers targeting the same window experience increased queuing delay, extending the critical path of the collective.

The irregular communication patterns illustrated in the hugetrace-00020 heatmaps (Figure~\ref{hugetrace-00020}) confirm this behavior. The sendcount and recvcount distributions are skewed, with certain ranks (e.g., ranks 5–7) receiving substantially more data than others. Under such imbalance, lock-based approaches are prone to lock contention, where concurrent requests to the same target form lock queues, causing waits and retries that inflate synchronization latency.

Fence-based persistent algorithms avoid this issue by relying on a single global synchronization to enter and exit the RMA epoch, eliminating per-target locks entirely. Consequently, Fence persistence is less sensitive to communication imbalance and provides more stable performance across irregular workloads.

%A useful analogy is to imagine a house with one door and four friends arriving: if all four 
%enter at once (global synchronization), the door opens and closes only a single time. If they 
%arrive one by one (per-target locking), the door must repeatedly open and close, and delays 
%accumulate while one person waits for another to finish. Similarly, 
%\texttt{winfence\_persistent} minimizes synchronization overhead, whereas 
%\texttt{winlock\_persistent} incurs locking delays, especially for busier target ranks.

\section{Conclusion}
\label{Conclusion}

This work designed, evaluated our persistent RMA implementations of 
\texttt{Alltoallv} using fence- and lock-based synchronization, separating 
one time initialization from per iteration execution. By caching window state, 
remote displacements, datatype decoding, and communication schedules in a reusable 
request object, the persistent path avoids repeated metadata processing and window 
setup, particularly in iterative workloads. 
A simple break-even analysis,  
\(
N_{\text{breakeven}}=\left\lceil
\frac{T_{\text{init}}}{T_{\text{MPI}}-T_{\text{persist}}}
\right\rceil,
\)  
formalizes when persistence becomes advantageous.

On our system, the Fence persistent variant consistently outperformed the 
non-persistent baseline for large message sizes and scaled more robustly with 
increasing process counts. For messages \(\ge 32{,}768\)~bytes, both Fence and 
Lock persistent achieved immediate payoff (\(N=1\)), with Fence reducing per-iteration 
runtime by up to 44\% and Lock by up to 38\% at \(2{,}097{,}152\)~bytes. 
For sizes \(\le 16{,}384\)~bytes, our Fence and Lock persistence variants did not provide a performance benefit, since the metadata costs eliminated by persistence were outweighed by the synchronization overheads inherent in the Fence- and Lock-based designs.

Under irregular sparse communication, similar trends were observed, confirming that 
Fence-based persistence is more reliable and less sensitive to synchronization costs 
than the lock-based approach.

\paragraph{Implications and future work}
For applications with repeated \texttt{Alltoallv} phases, Fence-persistent RMA
is the preferred choice for \(\ge 32{,}768\)~bytes (immediate payoff) and
 Future work
includes: (i) improving passive-target progress to reduce lock-epoch overheads,
(ii) integrating an adaptive runtime that selects between Fence-persistent and
two-sided collectives using measured \(N_{\text{breakeven}}\), and (iii)
extending the persistent RMA approach to other collectives.

\section{Acknowledgment}

I would like to acknowledge funding from the National Science Foundation under grant \# 2412182.  I also, acknowledge funding from Tennessee Technological University, the University of Tennessee at Chattannooga (SimCenter), and additional support from the Department of Energy through an internship at the Lawrence Livermore National Laboratory.

\bibliographystyle{ACM-Reference-Format}
\bibliography{ref}

\end{document}